# Co-Creating Neurorehabilitation Technology Alongside Lived Experience Experts: A Participatory Design Workshop

Irene L. Y. Beck[1,2]*, Salvatore L. Cucinella[1,2,3]*, Johannes B. J. Bussmann[1], Natacha Léon[4,5], José López-Sánchez[6], Francesca Lunardini[4,5], Katherine L. Poggensee[1,2,7], Cristina Vázquez-González[6], Joost de Winter[2], and Laura Marchal-Crespo

*These authors contributed equally

[1]Department of Rehabilitation Medicine, Erasmus MC, Rotterdam, The Netherlands

[2]Department of Cognitive Robotics, TU Delft, Delft, The Netherlands

[3]Department of Rehabilitation Medicine, Rijndam Rehabilitation Centre, Rotterdam, The Netherlands

[4]Hospital Los Madroños, Madrid, Spain

[5]Universidad Carlos III de Madrid, Madrid, Spain

[6]Centro Europeo de Neurociencias CEN, Madrid, Spain

[7]Department of Mechanical Engineering, University of Delaware, Newark, DE, United States of America

## Abstract

**Background:** Integrating the lived experience (LE) of patients early in the design process of neurorehabilitation technology can make the resulting solutions more relevant and effective. However, the incorporation of LE experts is often limited to testing finished prototypes.

**Objective:** This study aimed to present and evaluate a participatory design workshop for the early involvement of LE experts in neurorehabilitation technology development and to identify how their contributions influence design processes and outcomes. We further provide recommendations for researchers seeking to adopt similar approaches.

**Methods:** We conducted a participatory design workshop at the 6th International Conference on NeuroRehabilitation (ICNR 2024). Five people who have had a stroke (LE experts) and 18 workshop attendees (including the LE experts' caregivers) with experience in the field of neurorehabilitation were organized into five groups. Using the Double Diamond model, groups advanced through the Discover, Define, and Develop phases, guided by custom tools such as Journey Maps and Data Collection Notebooks, to co-create a solution addressing an objective specified by the LE expert. We applied thematic analysis of the Journey Maps and structured extraction of the audio transcripts of the groups' communication using a large language model.
**Results:** In all groups, the analyses indicated that the LE expert's input shifted the discussion from attendees' discipline-based assumptions toward priorities rooted in lived experience, such as fear, confidence, sensory discomfort, fatigue, aesthetics, and independence. Groups addressing narrowly specified objectives, such as opening a blister pack, designed focused, task-specific solutions. Conversely, broader objectives led to device concepts, including a sensory feedback system for cycling and a stimulation-equipped splint for cooking. Workshop satisfaction was high among attendees and LE experts. We share all workshop tools and protocols to support the adoption of the methods.
**Conclusions:** The proposed workshop methodology provides a practical foundation for early LE incorporation in neurorehabilitation technology development. Future iterations should better capture LEs' psychological and social dimensions, allow more time for reflection, and extend participatory design to the full range of relevant stakeholders. Achieving this requires sustained transdisciplinary engagement between researchers and the LE community.

## 1. Introduction

Over the past few decades, neurorehabilitation has undergone a technological revolution, impacting how patients are diagnosed and treated [1]. The growing interest in robotic devices, assistive mechanisms, and wearable systems is driven by an aging society, a higher incidence of associated neurological conditions, such as stroke [2], and increasing pressure on the healthcare workforce [3,4]. These innovations support therapists in providing high-intensity, repetitive, and task-specific training while lowering their physical effort [5]. They are often complemented with adaptive and gamified exergames to promote patient adherence and engagement [6,7]. Beyond clinical settings, assistive and portable devices are also increasingly found in patients' homes, enabling blended care neurorehabilitation [8] and support for activities of daily living (ADLs) [9].

However, despite the push from rehabilitation engineering researchers and developers, the adoption of neurorehabilitation technologies into widespread clinical practice remains limited. Even when shown to be effective, new approaches are slow to reach routine clinical care, with a frequently cited estimate of seventeen years to move from evidence to practice [10]. For example, in a recent questionnaire among 423 Italian rehabilitation professionals, 40% reported being unfamiliar with any robotic devices. Respondents reported financial barriers, such as limited resources and training, as well as barriers related to distrust and the perceived high complexity of the technology [11].

One strategy to address these barriers would be to involve more stakeholders earlier in the development process. While clinical expertise from therapists provides meaningful insights into how these technologies could be integrated into clinical practice, people with "lived experience" (LE), i.e., the knowledge held by individuals who personally navigate a health condition and its consequences in daily life [12], could contribute unique perspectives towards developing more effective, usable, and sustainable neurorehabilitation technology [13]. Incorporating LE experts as active research partners not only promotes the development of clinically relevant and effective technology [14–18] but may also empower LE experts and enhance public trust in healthcare [19,20]. Incorporating LE expertise from the inception of the project could enrich the entire design lifecycle, enhancing problem identification, research prioritization, participant recruitment, and community outreach. Yet, their involvement is often confined to advisory or consultative roles, such as assessing the usability of solutions after development [21], despite their desire to participate more actively in the development and decision-making process [22]. Therefore, researchers remain the decision-makers, while only few have LE experience, as researchers with LE of disability are themselves underrepresented and receive funding at lower rates than their non-disabled peers [23].

To answer this need, neurorehabilitation engineering could therefore benefit from human-centered design approaches, such as user-centered design and participatory design to more actively involve LE experts. While both frameworks share a commitment to engaging users, they differ in the timing and depth of involvement. User-centered design focuses on product usability, often involving users later in the process to test and refine designs, without contributing to the early definition of the research agenda (e.g., [24]) [25]. In contrast, participatory design, i.e., a form of co-creation applied across a design project, is rooted in the active involvement and shared decision-making of people with LE and other stakeholders throughout all project stages, as true partners [25]. By incorporating LE experts and those closely related to the experience, such as family members and therapists (referred to collectively as caregivers), in a co-creation process, engineers, designers, and developers can familiarize themselves with and empathize with the daily challenges they face and work together towards a solution that is satisfactory to all parties [26].

Participatory design and co-creation are increasingly adopted methods in neurorehabilitation engineering. However, researchers with traditional engineering backgrounds often have limited training in LE incorporation practices and qualitative methods, making the collection and analysis of human-centered data a challenge. Educational resources are emerging, including university curricula in human-centered design [27,28], internship programs like American Student Placements in Rehabilitation Engineering [29], and online courses (e.g., North American Spinal Cord Injury Consortium's Spinal Cord Injury Research Advocacy Course [30], European Patients Academy on Therapeutic Innovation [31]) that train researchers and LE experts to collaborate. However, access remains uneven, and researchers motivated to engage in co-creation may lack institutional support or proximity to such programs. Furthermore, in the neurorehabilitation field, cognitive, communicative, or physical impairments may affect how LE experts can engage in research activities and may require methodological adaptation and support to meet participants' capabilities and needs.

In short, there is a need for transdisciplinary collaboration to ensure meaningful LE expertise incorporation in the development of neurorehabilitation technologies [32,33]. To bring participatory design and co-creation education directly to neurorehabilitation engineering professionals, we designed and delivered a hands-on workshop grounded in a "learning-by-doing" philosophy [34] at the 6th International Conference on NeuroRehabilitation (ICNR 2024) [35]. Through this workshop, we aimed to broaden neurorehabilitation engineering researchers' understanding of participatory design methodologies, provide practical recommendations for participatory design application, and empower researchers to co-create together with LE experts, as essential partners, more effective, clinically relevant, usable, and inclusive neurorehabilitation technologies.

In this paper, we describe the development of and findings from this participatory design workshop in which academics, industrial partners, therapists, and LE experts and their caregivers, collaborated in groups to co-create neurorehabilitation technology concepts addressing needs put forward by the LE experts themselves. We used the Double Diamond model, which divides the design process into four phases—Discover, Define, Develop, and Deliver [36]—and engaged in the first three phases during the workshop. For each phase, we provided participants with custom co-creation tools and rapid, low-fidelity prototyping tools [37]. In the Discover phase, workshop attendees explored the lived experiences, rehabilitation goals and challenges of the LE experts. They broke down a specific activity identified by the LE expert prior to the workshop into steps to better empathize with specific challenges the LE expert may experience while executing it (Exercise 1). In the Define phase, the LE experts joined the groups and, building on the initial results from Exercise 1, the groups co-defined the design problem of the identified activity (Exercise 2). Finally, in the Develop phase, groups co-created a conceptual solution to the identified design problem through drawing and writing and presented their solution to the other groups (Exercise 3).

To evaluate the workshop, we adopted a mixed-methods approach, through participant questionnaires and qualitative analysis of the workshop materials, outcomes, and discussions. To support analysis of the relatively large volume of audio data, recordings were automatically transcribed and translated, and a large language model (LLM) was used to extract data from the transcripts.

## 2. Methods

### 2.1 Participants

The workshop was conducted at University Hospital Los Madroños (HLM, Madrid, Spain) as part of the 6th International Conference on NeuroRehabilitation (ICNR 2024). The workshop included 23 participants: 18 workshop attendees (14F/4M, age = 36.1 [*SD* = 10.7] years) and 5 LE experts ([former] patients; 2F/3M, age = 62.0 [*SD* = 12.8] years, with age available for 4 of 5 LE experts). Workshop attendees were clinicians (including the five caregivers introduced below), academics, and industrial representatives in the field of neurorehabilitation. All participants provided voluntary, written informed consent prior to the start of the workshop. The study was approved by the Comité de Ética de la Investigación Clínica of the Hospital Universitario Severo Ochoa, Leganés, Spain.

LE experts were recruited by clinical personnel from HLM and the Centro Europeo de Neurociencias (CEN, Madrid, Spain) in the month prior to the workshop. LE experts were eligible if they were adults (≥18 years) with ischemic or hemorrhagic stroke, at least one-month post-event, with preserved vision, unilateral motor impairment, and sufficient cognitive and communicative abilities to follow instructions, express opinions, and provide informed consent. LE experts with severe medical or psychiatric conditions, behavioral disorders interfering with participation, or inability to consent were excluded.

Workshop attendees were organized into five groups (Groups A to E), each arranged around its own table. Group size was limited to a maximum of six participants to reduce the cognitive load on the LE experts. One LE expert and their caregiver were assigned to each group. The caregivers were among the 18 workshop attendees: they were clinical personnel from HLM or CEN, and had a direct clinical relationship with the LE expert. The caregivers actively participated in all workshop exercises and supported the LE expert in the discussion, providing additional insights regarding the LE expert's recovery process as needed.

One participant per group served as *facilitator*. Their role was to guide the process, encourage collaboration, and provide support (e.g., translate from/to Spanish to/from English). In two groups, the caregiver served as facilitators while in the other groups, the facilitators were affiliated with the research team.

Finally, three members of the research team acted as *moderators*. They explained the exercises, custom-designed tools, and the participants' roles. Additionally, they collected the completed custom-designed tools, consent forms, and questionnaires from each group and managed the time.

## 2.2 Protocol

The workshop took place in a room arranged to allow sufficient space to move around and work in groups. It included one table per group, provided with Post-its, pens, and printed custom-designed tools to guide the workshop participants through the different activities. Each table was equipped with an audio recording device to record group discussions for later analysis. The workshop lasted two hours and included three introductory presentations and three hands-on co-creation exercises.

### *2.2.1 Introductory presentations*

The moderators started by explaining the workshop objectives, the code of conduct (i.e., work together; listen; build on each other's ideas; and propose ideas that you may feel are unconventional, regardless of utility, importance, or feasibility), the principles of participatory design (i.e., collaboration, inclusivity, empowerment; see [38]), the custom-designed tools, and the timeline of the workshop.

In the second presentation, a physician from HLM and an occupational therapist from CEN introduced the LE experts to the workshop participants to help them start empathizing with the LE experts, who were not yet present in the room. In particular, they shared the information previously collected from each LE expert and summarized in the *Data Collection Notebook*, a tool developed by the research team to provide each group with an overview of the person they would be designing with (see Figure 1). This overview included demographic information, upper-limb motor ability (Fugl-Meyer Assessment Upper Extremity scale [39]), cognitive abilities (Montreal Cognitive Assessment [40]). Clinicians employed the Canadian Occupational Performance Measure (COPM) to identify daily activities important to LE experts, along with their perceived performance and satisfaction with these activities [41]. Either photos or links to videos of the LE experts demonstrating those activities were included.

Finally, a neuropsychologist gave a presentation on communicating with people with stroke, highlighting the importance of considering each individual's physical, cognitive, emotional, social, and communication needs. It included explanations on how difficulties with perception, attention, memory, movement, and language can affect daily functioning and interaction. Attendees were encouraged to manage environmental stimuli, allow patients enough time to respond, and use motivating activities connected to everyday life.

HOSPITAL
Los Madroños

DATA COLLECTION NOTEBOOK

**Participant ID:**
**Age:** [redacted]
**Sex:** [redacted]
**Date of episode:** [redacted]
**Type of stroke:** Ischemic | Healthy subject ☐
**Affected side:** Left
**Dominance:** Right | Dominance:

| Test | Final result | |
|---|---|---|
| Fugl Meyer Upper Extremity | 65/66 | |
| Montreal Cognitive Assessment (MOCA) | 28/30 | |
| Canadian Occupational Performance Measure (COPM) | **Objective:**<br>-Remove the pills from the blister pack<br>-Open the package of cookies | |
| | **Performance**<br>Average score: 3,5 | **Satisfaction**<br>Average score: 6,5 |

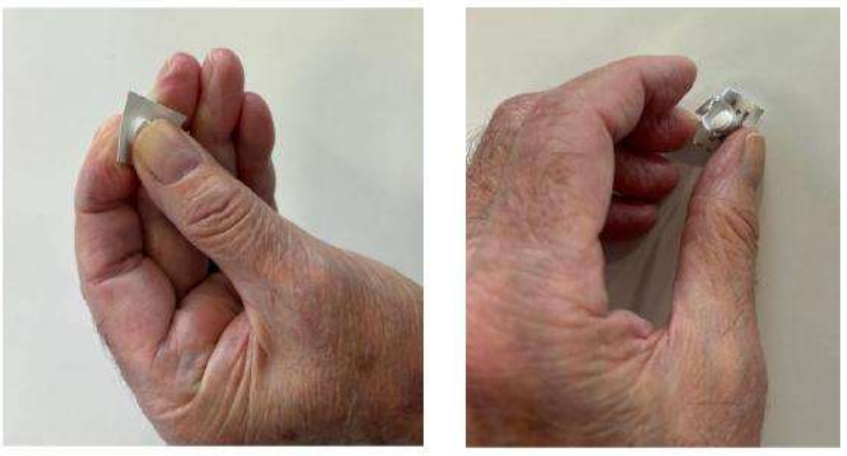

*Figure 1: Example of the Data Collection Notebook used by Group D during the workshop. The notebook, created together with the LE expert, includes their demographics and scores on several clinical tests. The performance and satisfaction scores for the Canadian Occupational Performance Measure are rated on a 10-point scale, with the LE expert illustrating one of the objectives. The Spanish version of the notebook was employed. Potentially identifiable information has been redacted.*

### 2.2.2 Hands-on research exercises

Following the oral presentations, we invited participants to engage in a series of hands-on exercises. They started with a 10-minute icebreaker within their group to promote self-expression, collaboration, and trust among participants [42]. Each participant filled in and shared a *Personal Profile Sheet* (Appendix A), a tool created by the research team, where they shared their name, profession, and answered questions such as "If you had to choose a different career, what would it be?" After the icebreaker, the three hands-on research exercises began, aligned with the first three phases of the Double Diamond: Discover, Define, and Develop [36].

#### Exercise 1 | Discover

Without the LE expert present, participants in each group collectively selected an objective from the *Data Collection Notebook* (e.g., "removing pills from a blister pack") to develop a *Journey* Map. A *Journey Map* is a design tool used to develop an empathetic understanding of end users by breaking down their goal activity into smaller steps and reflecting on each one [43,44] (Figure 2).

For each step, they could either draw or write the *Activity* and write down the *Goal of the Activity* (Figure 2). At the end of this exercise, workshop attendees were asked to complete a first questionnaire that captured basic demographics, prior experience with co-creation methods and interdisciplinary collaboration, and the perceived value of these approaches (see Section 2.3.1).

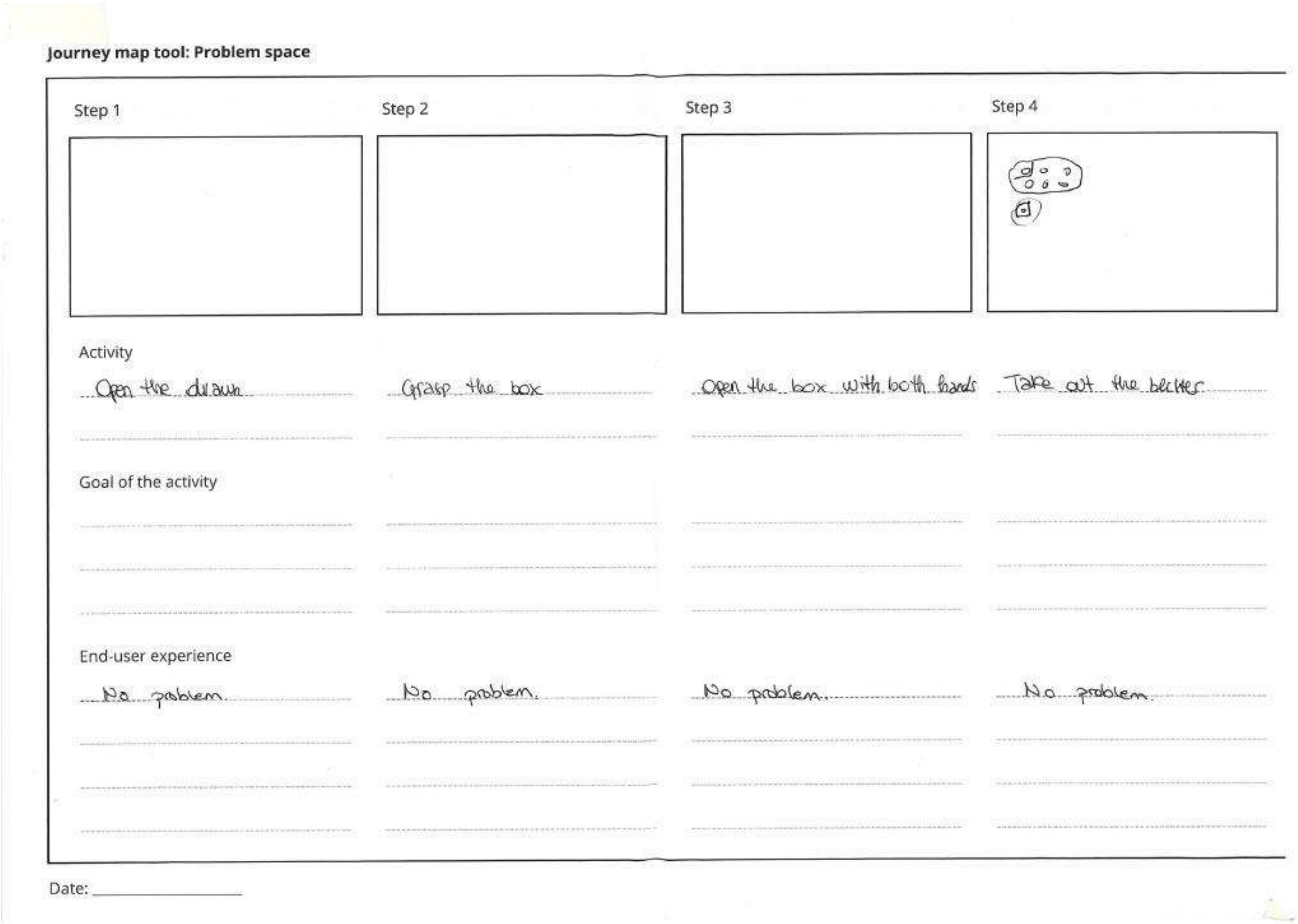

*Figure 2: Part of the Journey Map: Problem space tool filled by Group D. The tool was used by participants to break down an LE expert-defined goal (e.g., removing pills from a blister pack) into sequential steps by completing the Step, Activity, and Goal of the Activity sections (Exercise 1). The End-user Experience section was completed once the LE expert joined the group (Exercise 2).*

### Exercise 2 | Define

In the second research exercise, LE experts joined their assigned groups. Workshop attendees introduced themselves using the *Personal Profile Sheet* they had previously completed.

Workshop attendees, together with the LE experts, then refined and validated the *Journey Map* they had developed in Exercise 1 by incorporating the LE expert's perspective. Attendees were encouraged to interact with the LE expert, for example, by asking questions or observing explanatory movements, to better understand their experiences and needs in relation to the selected objective. Caregivers could support this process by helping the LE expert communicate, clarifying contextual details, or adding practical information about the selected activity.

During this exercise, attendees completed the *End-user Experience* section of the *Journey Map: Problem space tool* (Figure 2). This section was intended to capture the LE expert's input rather than attendees' assumptions. Based on the completed *Journey Map*, each group was then asked to co-define the core design problem, i.e., the aspect of the activity that is most problematic and in need of a design solution.

#### Exercise 3 | Develop

In the third research exercise, participants and LE experts engaged in an ideation session to co-create solutions addressing the design problem defined in Exercise 2. They were asked to express their ideas through discussion, drawings, and sketches, and to document them in the *Journey Map: Solution space* tool. This tool was designed to record the identified problem, the proposed solution, and a visual representation of the co-created concept (Figure 3).

Each group then presented their solution to the other groups in a two-minute pitch and received feedback. After this final exercise, workshop attendees completed a second questionnaire, and LE experts filled in a separate questionnaire to evaluate their experience in the workshop (see Section 2.3.1).

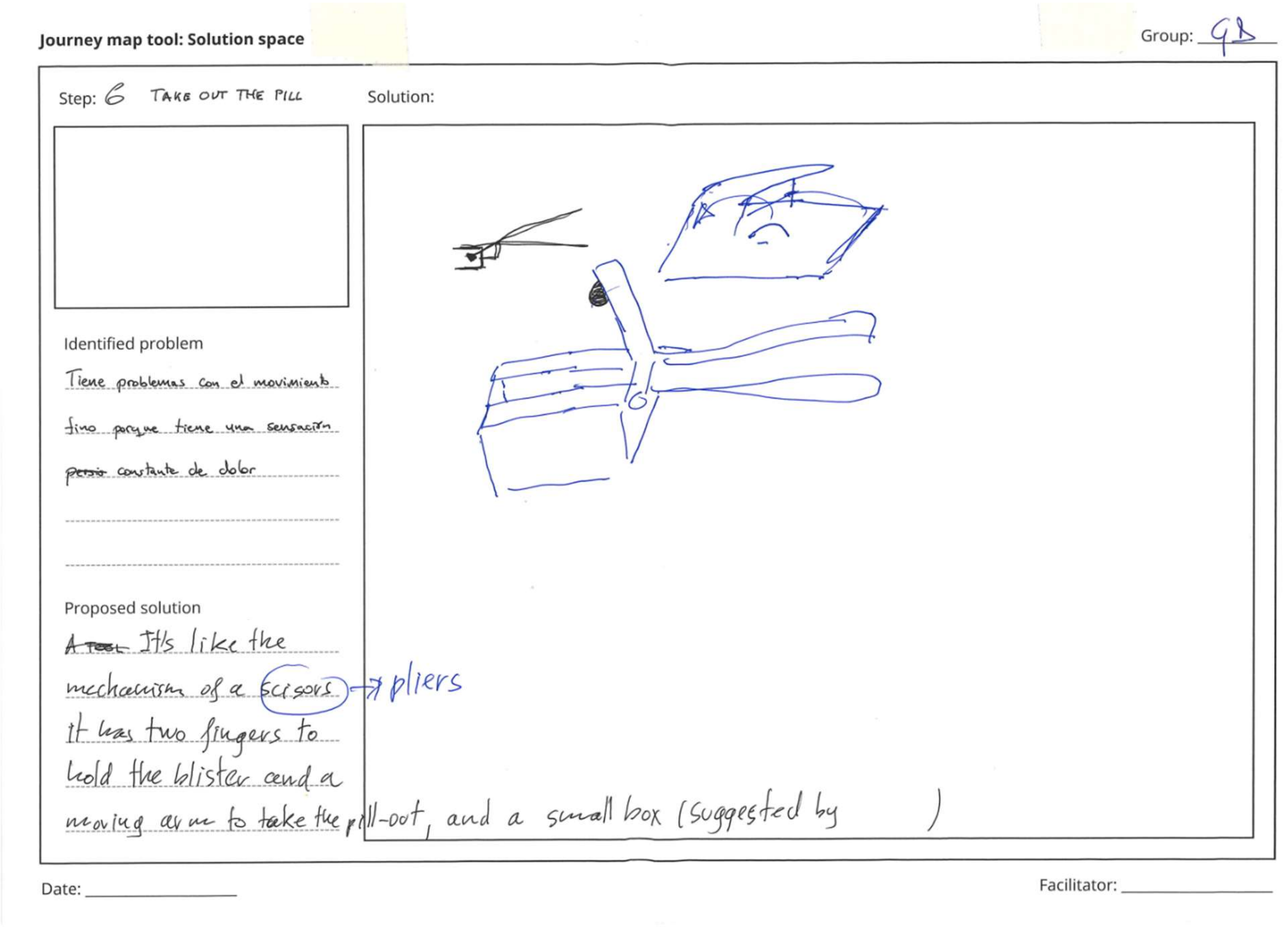

*Figure 3: Example of a completed Journey Map: Solution space tool, from Group D. Groups documented the co-defined problem with the LE expert, and outlined the proposed solution together with a sketch of the co-created idea. Potentially identifiable information has been redacted.*

## 2.3 Data collection and analysis

We collected both quantitative and qualitative data to evaluate participants' perceptions and experience of the co-creation process and to assess the strengths and weaknesses of the design methods and tools. We administered pre- and post-workshop questionnaires to workshop attendees and a separate post-workshop questionnaire to LE experts (Section 2.3.1). Additionally, we used an LLM to extract structured information from the de-identified, transcribed, and translated audio recordings of the group discussions (Section 2.3.2) and performed thematic analysis on the co-created Journey Maps (Section 2.3.3). The analysis of the qualitative data focused on the design process, the resulting solutions, and how the LE experts' contributions informed the design process and final solutions.

### *2.3.1 Questionnaires*

Workshop attendees completed two paper-based questionnaires. A first questionnaire (Tables 1 & 2) was administered after Exercise 1 and included demographic questions (e.g., gender, age, job title, field of work, years of experience), six Likert-scale items on prior experience, perceived value, and confidence regarding co-creation methods and interdisciplinary collaboration (1 = No experience / Do not value / Not confident to 5 = Expert level / Highly value / Very confident), and four statements on experience and enjoyment working with end-users and how such collaborations foster empathy and improve the design of novel technologies (1 = Not at all to 5 = Very much so).

A second questionnaire (Tables 2 & 3) was completed after the last exercise. Attendees reported their overall satisfaction with the workshop (10-point scale) and re-answered five of the six Likert-scale items from the first questionnaire. They also rated their agreement with 14 additional statements (1 = Not at all, 5 = Very much so) assessing their engagement and perceived contribution during the workshop, their intention to adopt the methods learned, and their attitudes toward end-user involvement in the design process. Seven open-ended questions addressed their overall experience, views on LE expert involvement, awareness of existing design methods, suggestions for future workshops, key takeaways, and desired follow-up resources. The seven open-ended questions are reported in Appendix B.

LE experts completed a separate post-workshop questionnaire after the last exercise (Table 4). They rated their overall experience (10-point scale) and their agreement with 13 statements (1 = Not at all, 5 = Very much so) evaluating whether they felt heard, understood, and valued; whether they felt actively engaged; and to what extent the design process and resulting solution reflected their needs, preferences, and input. Four open-ended questions addressed their overall experience, suggestions for improving the design process, whether the resulting solution met their needs, and whether participation made them feel more in control of their recovery journey (see Appendix B). The open-ended responses were translated into English by a native Spanish-speaking researcher.

#### *2.3.2 Audio recordings*

We used an LLM to support structured analysis of the audio recordings from each group. The model was asked to extract the final solution, the most important design considerations leading to that solution, and the considerations specifically raised by the LE expert in each group.

The audio recordings (approximately 1.5 hours per group) were transcribed locally using Whisper large-v3 (OpenAI, USA). A Spanish-speaking researcher de-identified the transcripts by redacting names and identifying references. The de-identified transcripts were translated from Spanish to English using DeepL (DeepL SE, Germany). A Spanish-speaking researcher verified the quality of the translations using the transcript of Group B.

We used Microsoft 365 Copilot, running GPT-5.5 in Think Deeper mode (OpenAI, USA) under the university work account, to extract structured information from the translated de-identified transcripts. Under Microsoft's enterprise data protection terms for Copilot Chat, prompts and responses are not used to train foundation models. Two authors initially iterated on the prompt using Group B's transcript. The following prompt was then used for each transcript:

```
1. What is the final product solution created by the group
during a co-creation workshop? And what were the 10 most
important considerations mentioned leading up to the final
design (maximum 5 words each), supporting each consideration
with a verbatim quote from the transcript (mention which part
this quote was taken from).
2. Lastly, what were the considerations brought forward by the
patient. Provide a list of maximum 5 considerations (maximum 5
words each) and support each consideration with a verbatim quote
from the transcript (mention which part this quote was taken
from).
```

To validate the quality of the LLM-extracted information, two authors independently performed the same extraction task, using the same prompt as instructions, on the transcripts from Group A and C (one group each), before reviewing the Copilot output. The human and LLM extractions are reported side-by-side in Appendix C. The human and LLM extractions overlapped on many considerations, although differed in their emphasis. We judged the agreement adequate for our purposes, and the LLM-assisted approach was subsequently used for the remaining three groups. All LLM-extracted quotes were verified verbatim against the transcripts. We synthesized the extracted information into a narrative summary organized around the final solutions, the role of LE expert input in shaping design considerations, and recurring patterns across groups.

### *2.3.3 Journey Maps*

The *Journey Maps* were thematically analyzed using the Analysis on a Wall approach [45,46], following an adaptation of the six phases of thematic analysis [47]. The analysis was conducted by four researchers from TU Delft and focused on the influence of LE expert involvement in the design process. Phases 1 through 5 were performed, with phases 2 (generating initial codes) and 3 (searching for themes) executed simultaneously, as well as phases 4 (reviewing themes) and 5 (defining and naming themes). Phase 6 (producing the report) corresponds to the writing of this paper.

Each researcher independently read and reviewed the *Journey Maps* displayed on the walls to familiarize themselves with the data (Phase 1). They then individually generated initial codes using Post-it notes placed next to each *Journey Map*, capturing observations related to the design process and the LE experts' influence on that process, and identified preliminary themes (Phases 2–3). The researchers then met to review and discuss their individually generated themes (Phases 4–5). For each *Journey Map*, the Post-it notes were clustered into final themes through discussion until consensus was reached. Finally, themes across all *Journey Maps* were compared and synthesized to define overarching themes.

# 3. Results

## 3.1 Questionnaires

### *3.1.1 Workshop attendees' responses*

Seventeen of the 18 workshop attendees completed the pre-workshop questionnaire, and 15 completed the post-workshop questionnaire. Attendees came from diverse professional backgrounds in neurorehabilitation and related disciplines, including robotics, neuroscience, biomechanics, bioengineering, and neurotechnology. Participants included clinical personnel (4 occupational therapists, 3 physiotherapists, a neuropsychologist, and a head of rehabilitation unit; $n$ = 9), researchers and academics (3 PhD students, a researcher, and an Associate professor; $n$ = 5), and engineers and technology developers (a rehabilitation engineer, an R&D specialist, and a sales manager (engineer); $n$ = 3). On average, workshop attendees were in their current position for 6.0 years ($SD$ = 6.6) and had 12.4 years of professional experience ($SD$ = 9.8).

Three co-authors participated as facilitators, and one as participant, and their questionnaires' responses were recorded. Their questionnaires' ratings are excluded from the analysis, while their answers to the open-ended questions are included. Appendix D shows the questionnaire scores' means and standard deviations including their data. Most means and standard deviations differed by at most 0.1 points on the 5-point scale. The attendees' responses to the pre-workshop statements (Table 1) indicated relatively high prior experience with interdisciplinary collaboration and with working with LE experts (referred to as "end-users"), an activity they reported as enjoyable. On average, attendees strongly agreed that interactions with LE experts foster empathy and understanding of their needs and challenges, and that these interactions improve the design of novel technologies and solutions.

Responses to the repeated questions (Table 2) showed that, before the workshop, attendees did not regard themselves as experienced in co-creation methods, despite highly valuing both co-creation methods and interdisciplinary collaboration. All scores' mean increased or remained stable after the workshop, with the largest increase related to confidence in applying co-creation methods (from 3.1 to 4.1).

Workshop attendees rated the workshop highly, with a mean response of 9.1 (*SD* = 1.2) to the question "Please rate the workshop on a scale of 1 (I didn't like it at all) to 10 (I liked it very much) based on how much you liked it." There was also a high level of agreement with the statements in the post-workshop questionnaire (Table 3). In general, attendees reported feeling actively engaged in the workshop (mean = 4.7, *SD* = 0.7). In line with the pre-workshop statement ratings, they agreed very much on the value of collaborating with, empathizing with, and understanding LE experts in the design process (means of 4.9, 5.0, and 5.0, *SD*s of 0.3, 0.0, and 0.0, respectively). Their mean scores concerning the significance of their own expertise to the group's success and the extent to which the workshop helped them identify strengths and weaknesses in their current design process were relatively lower (4.1 and 4.0). Finally, attendees indicated that they intend to apply the methods learned and to actively involve LE experts in future design projects (means of 4.2 and 4.8, respectively).

When asked what they liked most about the workshop, 13 attendees answered, 10 of whom cited the involvement of LE experts. They described activities such as asking questions, listening to, and collaborating with LE experts as especially valuable. Five attendees also mentioned the activity's multidisciplinary or interdisciplinary nature (*e.g., "Being with the patient in a multidisciplinary environment")*. When asked about the active involvement of LE experts in the design process, all 12 attendees who answered described it as important (e.g., *"Extremely necessary"*, *"It's a must"*). Attendees emphasized that direct interaction with LE experts kept the design process focused on their real needs: *"I liked that I could directly ask crucial questions from the end-user, therefore it was easier not to get lost in a side path; it was definitely more focused."* One (co-author) facilitator noted that the interaction was easier than expected: *"It is easier than I thought. They were very motivated to help. It made me think about other possibilities."* LE expert involvement was the most common key takeaway, including the importance of listening to emotions and experiences, incorporating patient ideas, and recognizing patient-specific needs: "*I liked working with different people with different experiences in order to be able to see different points of view, mostly the patient's, that's crucial for creating a solution*."

Time pressure was the main point of criticism. Ten of the 11 attendees who described their least-liked aspect reported the limited time, and 9 of the 11 who suggested improvements recommended allowing more time for discussion, reflection, and explanation of steps (e.g., *"It would have been better to have more time and discuss more objectives, but I understand it is difficult"*). The only other least-liked response concerned language: *"The* *need to translate broke up the dynamics of the method."* The remaining suggestions concerned greater diversity in participant backgrounds and preparatory materials, such as videos of patients performing daily activities, a suggestion made by a participant whose group did not have a video attached to the *Data Collection Notebook*.

Finally, in response to the question about follow-up activities or supporting resources to help implement the workshop's teaching more effectively, attendees listed additional materials, such as online resources, publications, videos, and presentations. This seems essential based on the limited previous experience with design methods; in a separate question, only five of eight participants who answered provided specific known/applied methods (e.g., iterative prototyping and user testing, analytic hierarchy process, experience diaries, questionnaires), one referred to psychological approaches in general terms, and two reported no such experience. Some proposed further collaboration with patients as follow-up activities: *"[It] would be nice to work on different patients and maybe try the solution together with them."* Yet, as noted by another participant in a different question: "*maybe I see an applicability issue due to scarce interaction with patients in early stages of research."*

Table 1. Workshop attendees' responses to five pre-workshop items. [a]

| Statement | Pre-workshop score Mean (SD) |
|---|---|
| | |
| How much experience do you have with interdisciplinary collaboration? (1 = No experience, 5 = Expert level) | 4.6 (0.8) |
| I have experience working with end-users (e.g., patients and caregivers). (1 = Not at all and 5 = Very much so) | 4.1 (1.2) |
| Interactions with end-users (e.g., patients and caregivers), both personally and professionally, can improve my approach to designing novel technologies and solutions. (1 = Not at all and 5 = Very much so) | 4.5 (0.7) |
| Working with end-users (e.g., patients and caregivers) allows me to better empathize with them and increase my understanding of their needs and challenges. (1 = Not at all and 5 = Very much so) | 4.8 (0.4) |
| I enjoy working with end-users (e.g., patients and caregivers). (1 = Not at all and 5 = Very much so) | 4.5 (1.1) |

[a]*Note.* $n$ = 11 for the interdisciplinary-experience item and $n$ = 13 for the four end-user statements.

Table 2. Workshop attendees' responses to the five questions administered both pre- and post-workshop. Anchors per question are shown in the table. [a]

| Question | Pre-workshop score Mean (SD) | Post-workshop score Mean (SD) |
|---|---|---|
| | | |
| How much experience do you have with co-creation workshops?<br>(1 = No experience, 5 = Expert level) | 1.7 (1.2) | 2.3 (1.4) |
| How much do you value co-creation methods?<br>(1 = Do not value, 5 = Highly value) | 4.2 (1.0) | 4.6 (0.5) |
| How confident are you in the application of co-creation methods?<br>(1 = Not confident, 5 = Very confident) | 3.1 (1.2) | 4.1 (1.0) |
| How much do you value interdisciplinary collaboration?<br>(1 = Do not value, 5 = Highly value) | 4.8 (0.6) | 4.8 (0.4) |
| How confident are you in your ability to collaborate across disciplines?<br>(1 = Not confident, 5 = Very confident) | 4.0 (1.1) | 4.3 (1.0) |

[a]*Note.* Pre-workshop $n$ = 11 for the first four items and $n$ = 13 for the fifth; post-workshop $n$ = 12.

Table 3. Workshop attendees' agreement with statements post-workshop. 1 = Not at all, 5 = Very much so. [a]

| Statement | Post-workshop score Mean (SD) |
|---|---|
| | |
| I felt actively engaged in this workshop. | 4.7 (0.7) |
| My expertise contributed significantly to the success of the group. | 4.1 (0.8) |
| I am likely to apply the methods and tools learned in this workshop to future design projects. | 4.2 (0.7) |
| This workshop helped me identify strengths and weaknesses in my current design process. | 4.0 (1.1) |
| End-users (e.g., patients and caregivers) participation enhanced the design process. | 4.9 (0.3) |
| Collaborating with end-users (e.g., patients and caregivers) was a valuable experience. | 4.9 (0.3) |
| Empathizing with end-users (e.g., patients and caregivers) is essential for designing solutions that meet their needs. | 5.0 (0.0) |
| Understanding end-users (e.g., patients and caregivers) needs is crucial for defining design requirements. | 5.0 (0.0) |
| Working directly with end-users (e.g., patients and caregivers) could be effective for understanding their needs. | 4.9 (0.3) |
| The involvement of end-users (e.g., patients and caregivers) made me reconsider my approach to design. | 4.4 (0.9) |
| Working with end-users (e.g., patients and caregivers) allowed me to better empathize with them and increased my understanding of their needs and challenges. | 4.9 (0.3) |
| I plan to actively involve end-users (e.g., patients and caregivers) in future design projects. | 4.8 (0.5) |
| Interactions with end-users (e.g., patients and caregivers) improved my approach to novel technologies and solutions. | 4.6 (0.7) |
| I believe co-creation could be an effective method for designing meaningful solutions. | 4.7 (0.5) |

[a]*Note.* $n = 12$.

### *3.1.2 LE experts' responses*

The LE experts rated the workshop highly, with a mean score of 9.6 (*SD* = 0.9) to the question "Please rate the workshop on a scale of 1 (I didn't like it at all) to 10 (I liked it very much) based on how much you liked it." They reported high levels of engagement (mean = 4.8, *SD* = 0.4) and feeling seen, safe, and understood (all means = 5.0; Table 4). They also felt their participation was meaningful (mean = 4.8, *SD* = 0.4), and the design process and final solution captured and reflected their needs (both means = 5.0, *SD*s = 0.0). They also reported feeling more empowered in their recovery process and increased confidence in using technology (both means = 4.4, *SD*s = 0.9).

The responses to the four open questions support these ratings. When asked about their overall experience as active participants in the design process and how their participation contributed to the design, four of the five LE experts highlighted the opportunity to share their personal experiences and to be heard: *"Expressing myself with complete sincerity and feeling heard and understood;" "The collaboration between industry and users makes us feel involved in the design of solutions."*

Evaluations of the resulting solution and of their sense of control over their recovery journey (three respondents per question) were predominantly positive, although one LE expert gave a mixed answer regarding control: "*On one side yes, because my perception* *can help, but on the other hand not because I think it is very hard to make it happen"*. Two LE experts indicated that the solution could support their daily life (*"I think it's a very good solution that would make my daily life easier"*), and two described an increased sense of self-efficacy and active contribution (*"It really does make me feel like it's making my recovery process easier because it solves a problem with my independence—I don't have to depend on anyone"*). The only critical response concerned the road from concept to implementation rather than the workshop itself (*"The problem is always to go from the idea to the implementation…"*), while expressing optimism about advancing technology and growing awareness of disability-related issues.

When asked which aspects of the design process worked well and what could be improved, two LE experts responded: one felt the process was well adjusted to their needs and difficulties, and one noted that the workshop's focus on a specific objective left other relevant aspects unexplored, although they appreciated hearing the solutions developed by the other groups.

Table 4. LE experts' agreement with post-workshop statements. 1 = Not at all, 5 = Very much so. [a]

| Statement | Post-workshop score Mean (SD) |
|---|---|
| | |
| I felt seen, heard, and acknowledged throughout the workshop. | 5.0 (0.0) |
| I felt safe and comfortable during the workshop | 5.0 (0.0) |
| I felt understood by the facilitators and workshop attendees during the workshop. | 5.0 (0.0) |
| I felt actively engaged and involved in the design process throughout the workshop. | 4.8 (0.4) |
| My participation in the design process felt relevant and meaningful. | 4.8 (0.4) |
| I enjoyed the collaborative aspects of the design process during the workshop. | 4.6 (0.9) |
| The questions asked during the workshop were relevant to my experiences and needs. | 4.8 (0.4) |
| My group seriously considered my opinions and feedback. | 5.0 (0.0) |
| The design process effectively captured my personal needs. | 5.0 (0.0) |
| The resulting solution reflects my needs, preferences, and input. | 5.0 (0.0) |
| Being involved in the design process made me feel more empowered in managing my recovery journey. | 4.4 (0.9) |
| My involvement in the design process increased my confidence in using the technology. | 4.4 (0.9) |
| I am satisfied with the level of involvement I had in shaping the resulting solution. | 5.0 (0.0) |

[a]*Note. n* = 5.

## 3.2 Audio recordings

For each group, the LLM extracted a description of the final proposed solution, the ten most important considerations leading up to that solution, and five considerations raised by the LE expert. Below, we summarize this output per group, illustrated with quotes extracted by the same model; the full LLM-generated output is reported in Appendix C, side by side with the human analysis outcomes for Groups A and C. The LLM-extracted design solution is summarized in Table 5.

Table 5. Summary of each LE expert's objective from the Data Collection Notebook, the resulting problem specification and outcome specificity identified through thematic analysis of the Journey Maps, and the co-created solution identified from the LLM-extracted audio-recorded group discussions.

| Group | LE expert's objective | Problem specification after Define phase | Outcome specificity | Co-created solution |
|---|---|---|---|---|
| | | | | |
| A | Riding a bicycle | Broad: fear of falling, low confidence, reduced proprioception | Concept device supporting a broader task | Sensory feedback system signaling loss of balance or limb misalignment while cycling, to support relearning of anticipatory postural control |
| B | Showering independently | Broad: temperature hypersensitivity, standing, neglected side, fixed object placement, independence | Task-specific solution | Shower-assist device combining body support with a rotating washing mechanism for the |
| | | | | neglected side of the body |
| C | Cooking with the affected arm | Broad: grasp support, wrist stabilization; unobtrusiveness and aesthetics across activities | Concept device supporting a broader task | Wearable wrist/hand splint with integrated stimulation to support grasping and stabilize the wrist |
| D | Removing pills from a blister pack | Narrow: painful fingertip pinch, pill falling on release | Task-specific solution | Pliers-like blister-pack opener operated with whole-hand pressure, with an attached box to catch the pill |
| E | Clipping fingernails | Narrow: one-handed stabilizing and pressing, visual control, fatigue | Task-specific solution | Table-mounted one-handed nail clipper with adjustable positioning and an extended press-down lever |

Groups D and E, whose LE experts defined narrow objectives, and Group B each converged on task-specific product solutions. Group D, working on opening a pill blister pack, designed a pliers-like opener that converts the task from painful fine fingertip movements (“*the strength to squeeze here, at the tips of the fingers, here with the fingers—I do have some differences there*”) into gross hand movements: “*we came up with a kind of pliers where you can place the blisters, and then, using the whole hand, something pushes the blister and removes the pill.*” The LE expert additionally suggested “*securing a box in the pliers to hold the pill in place while pressing,*” preventing the released pill from falling. Group E, whose attendees noted that the LE expert “*can cut her left fingernails, but she can’t do it on her right hand,*” designed a one-handed nail clipper mounted on a stable table platform (“*placing it in a stable position so that you don’t need to—maybe you don’t need to hold it*”), with adjustable positioning and an extended press-down lever to retain visual control: “*it might be better to see what you’re doing and have something that can be precise.*”

Groups A and C, working from broader objectives, initially focused on biomechanical and motor control aspects until the LE experts redirected the discussion. The resulting solutions in both groups were concept devices supporting the broader task. In Group A, where the objective was riding a bicycle, the LE expert explained that “*balance scares me the most*” and described the difficulty of “*coordinating, grabbing the handlebars, coordinating, moving my leg, and at the same time trying not to fall*.” As the LE expert had reduced proprioception on one side of the body, the group proposed a bike-based sensory feedback system that signals when balance, limb position, or alignment is off, intended to support relearning rather than to compensate: “*whether what we need is a motor that readjusts the bike or something that signals him to do it, because what we want is for him to learn how to do it.*” The LE expert in Group C had a similarly broad objective (“*the goal then was to cook with the left arm*”), but prioritized unobtrusiveness (“*my goal with the arm, even though we have this exercise, is for it to bother me as little as possible*”) and aesthetics (“*I customize them and design them to match my outfit*”). The final concept was a wearable wrist-hand splint with integrated stimulation to support grasping and to stabilize the wrist during cooking and other daily activities.

Group B achieved a concrete product despite its broad objective. The attendees initially mapped the shower routine in functional terms (“*it’s like the whole journey… to go into the shower, take a shower, get dressed, and then leave*”), whereas the LE expert emphasized hypersensitivity to temperature changes (“*For me, showering is a very unpleasant process. Because of hypersensitivity. To the temperature*”), standing instead of sitting, and independence: “*I shower, all by myself. I don’t let [redacted] take over—I do it myself.*” The final concept was a shower-assist device combining body support with a rotating washing mechanism that cleans the neglected body’s side while the LE expert leans or holds for balance.

Across the five groups, LE experts shifted attention from what attendees assumed was difficult toward what they experienced as difficult in daily life: fear of falling and loss of confidence in Group A, sensory hypersensitivity in Group B, grip and wearability in Group C, pain and weak fingertip pinch in Group D, and step order, precision, and fatigue in Group E.

## 3.3 Journey Maps

Three overarching *themes* emerged from the thematic analysis of the *Journey Maps*: (1) Specificity of the objective defined by the LE expert, (2) LE expert involvement, and (3) Utilization of the Journey Map.

### *3.3.1 Specificity of the objective defined by the LE expert*

The specificity of the objective noted in the *Data Collection Notebook* was associated with differences in the group discussions’ focus and dynamics. The *Journey Map*s associated with higher-level objectives were split into broader steps. For example, the generic objective “riding a bicycle” in Group A was broken down into balance, motor control, and muscle strength components during Exercise 1. Once the LE expert joined, they identified confidence as a central challenge, reflecting how the lived experience shifted the discussion's focus. Groups starting with a more specific objective broke it down into narrower steps. For example, two of the steps of the specific objective set in Group E, "clipping fingernails," were "stabilizing the nail cutter" and "positioning the fingernail inside the cutter." LE experts validated certain steps as "easy" or "no problem here" and directed attention to the steps they found most problematic. A corresponding pattern was observed in the specificity of the final design outcomes (Table 5): groups whose design problem was narrowly specified after the Define phase produced a product solution for a dedicated task, including Group B, whose LE expert narrowed a generic objective into concrete constraints, whereas Groups A and C developed concept devices supporting broader tasks.

### *3.3.2 LE involvement*

During Exercise 1 without the LE experts, the groups relied on their professional expertise to break down the objective. They focused on the motor function and the physical task, using terms such as "*absence of force in flexor/extensors*" or "*coordination of lower limb*" and "*take out the pill*."

The LE experts' input on the *Journey Maps* was recorded in the *End-user Experience* subsection during the following exercises, with explanations/comments added for each step. These included reasons why a step was challenging (e.g., "*She says she has difficulties doing two things at the same time"*) or that they were able to perform that step (e.g., "*there is no problem*"). Their experiences were more nuanced than the attendees' step descriptions from Exercise 1 and provided contextual information about their abilities to perform a task, reflected in the J*ourney Map: Solution space* tool. Those notes were consistent with the extracted information from the audio-recordings.

### *3.3.3 Utilization of the Journey Map*

The *Journey Map: Problem space* included four sections for each step of an activity: (1) a drawing space for the activity, (2) a text space for the activity, (3) a text space for the goal of the activity, and (4) a text space for the end-user experience. We found that each group approached the *Journey Map* differently.

Most groups used the drawing space to draw; however, some wrote inside it instead or did not use it at all, writing only in the text box. Overall, of the three sections available during Exercise 1, most groups used only one or two. The use of the *Journey Map* also varied across the defined steps: one group used all the drawing boxes to write the general text but added a detailed description and goal of the activity only for the first two steps. These variations indicate that the template was taken up differently across groups and that not all sections were completed within the available time.

# 4. Discussion

Co-creating neurorehabilitation technology within transdisciplinary settings requires methods that meaningfully embed LE experts from the earliest design stages. We introduced a set of participatory design methods and tools for early involvement, structured around the Double Diamond model. We used a mixed-methods approach to evaluate the co-creation methodology, custom-developed tools, and the influence of LE experts' participation on design considerations during an international conference workshop. To support the analysis of extensive audio recordings, we used an LLM as part of the evaluation process.

## 4.1 The active involvement of the LE experts reoriented group discussion and informed design outcomes

Involving LE experts early in the design process uncovered nuanced insights into their priorities and challenges. Their input went beyond feedback on proposed solutions; LE experts helped the groups identify the core design problems and shape the design process towards solutions aligned with the needs and desires they expressed.

Before the workshop, the *Data Collection Notebook* documented information typically used to define LE experts' clinical needs, namely, their levels of sensorimotor and cognitive abilities and the specific tasks they aim to accomplish throughout their neurorehabilitation journey. Combining clinical assessments with LE experts' reported priorities allowed workshop attendees to begin exploring their needs before meeting them. However, direct LE involvement revealed that the *Data Collection Notebook* alone did not provide a holistic view of the individual; important challenges, including psychological factors such as confidence, self-efficacy, aesthetics, and independence in performing tasks, were not sufficiently captured. This is consistent with the model of *Person-centered care* [48], which emphasizes understanding LE experts' experiences, preferences, and needs to tailor care to the individual. Future versions of the *Journey Map* could explicitly address the psychological and social dimensions of LE alongside functional challenges, thereby stimulating a deeper understanding of LE experts' needs and desires as early as possible.

Attendees similarly reported that LE experts' input helped keep the discussion focused and prevented groups from getting sidetracked. This was especially important when the objectives set out by the LE expert in the *Data Collection Notebook* were rather broad (e.g., “riding a bicycle”) compared to more specific goals (e.g., “removing pills from a blister pack”). Possibly due to the workshop's limited duration, broader objectives led groups toward concepts that supported an overall task, whereas objectives that could be narrowed into concrete constraints for a dedicated task led to more concrete product solutions. These findings indicate that broad vs. specific objectives may lead groups toward different types of design outcomes, and that broader objectives may require additional steps of problem definition before they can be translated into specific solutions.

This raises the question of whether it is preferable to start with a specific or a broad objective. As in reinforcement learning, there is a trade-off between exploration and exploitation: broader objectives supported exploration of the problem space (i.e., groups spent much of the session identifying which aspect of the activity was most problematic), while more specific objectives supported exploitation toward a more concrete solution (i.e., groups could proceed directly to solution development). This mirrors the first diamond, which combines divergent exploration in Discover with convergence on a defined problem in Define [36].

Exploration could also be facilitated by involving more LE experts per group, capturing a larger set of needs, priorities, and perspectives. As one LE expert stated: "*we focused on a specific aspect, but there are a lot of aspects that are also valid.*" Individuals with disabilities are diverse [49], and a solution shaped around one LE expert's abilities, goals, and context will not necessarily work for another without adapting it. This is typical of participatory design, where individual lived experiences lead to solutions tailored to specific needs and contexts, when they act as experts of their own experience rather than representatives of a broader population [25]. If the aim is to understand how a solution can be adapted to different people and contexts, future work should involve multiple LE experts across iterative design cycles, while still preserving the value of individual lived experiences. For the process to then converge toward a concrete solution, insights generated from multiple LE experts can be synthesized into shared design targets [46], for example, through the development of *personas*: fictional user representations that combine common patterns across participants while still reflecting each individual's needs, experiences, behaviors, and goals [50]. This choice can be made together with the LE experts and the project team, in line with the shared decision-making inherent in participatory design [25].

The workshop's design outcomes primarily addressed independence in ADLs rather than reducing the underlying motor or cognitive impairments. Each LE expert's objective was defined using the COPM and therefore inherently focused on meaningful ADLs. Although workshop attendees initially used their expertise and knowledge to address this objective from a motor function perspective, LE experts redirected the attention toward their need for independence (*"it solves a problem with my independence—I don't have to depend on anyone"*). Impairment reduction and improved activity and participation represent fundamentally different yet complementary objectives [51]. Technologies targeting impairment reduction focus on restoration of motor, sensory, and cognitive capacities, whereas those targeting activities and participation focus on improving performance, independence, and quality of life. While rehabilitation approaches may lead to more sustained independence, the path to reach the set goals can be slow. Furthermore, assisted repetitive movements might also result in improved function. This may explain why, in the short timeframe of our workshop, the groups gravitated towards assistive solutions.

For technologies aimed at activities and participation, LE experts have unique knowledge to identify daily-life challenges, which argues for a central role from the earliest stages of problem definition as equal members of the team. In contrast, when the goal is to design technologies aimed at restoring motor, sensory, and/or cognitive capacities, the initial problem identification requires additional expertise from clinicians [21,52,53]. Nevertheless, early LE involvement is crucial to inform which outcomes are meaningful to them and provide input based on their needs, daily experiences, and the technology context of use, e.g., when the technology is used at home without direct clinical supervision. The purpose of technology, therefore, determines not whether LE experts should participate, but how their expertise complements that of clinicians and other stakeholders.

## 4.2 A meaningful experience for both workshop attendees and LE experts

Overall, workshop attendees valued how transdisciplinary collaboration with LE experts helped them identify and understand the LE experts' needs. LE experts reported feeling heard and understood, being satisfied with their level of involvement, and seeing their preferences and needs reflected in the ensuing solution. The icebreaker using the *Personal Profile Sheet* may have supported this experience by initiating conversations and helping participants understand one another's backgrounds and perspectives, thereby facilitating an open and inclusive environment [25,42]. A safe and inclusive environment was further facilitated by the small-group format, the presence of a familiar caregiver, and the opportunity for LE experts to take breaks from the activities, as necessary [46,54].

LE experts also reported that their involvement increased their confidence in using the technology, a predictor of technology adoption [55,56]. The reported high levels of empowerment extend beyond research quality alone. The United Nations Convention on the Rights of Persons with Disabilities establishes the right of individuals with disabilities to participate in decision-making processes that concern them [57,58]. Participatory approaches offer one way to realize this right in the development of technologies that affect their lives [22].

## 4.3 Large Language Models to support the analysis of qualitative data

The role of LLMs in qualitative research is debated, with researchers pointing to the limited capabilities of such models in interpreting qualitative data and generating meaning from it [59], while others report that LLM annotations can match or exceed those of human coders [60]. Recently, Hill et al. [61] compared ChatGPT-5, Claude 4 Sonnet, and QualiGPT against blinded human analysts on healthcare focus-group data and found LLMs to be non-inferior to humans for deductive coding, but more variable in inductive analysis requiring interpretation of tone and context. Strict hallucination rates were low (1.2%), but when partial matches and incorrectly coded facilitator speech were included, error rates reached 12.4%, indicating that human verification remains necessary.

In this study, the use of automated transcription, translation, and LLM-assisted data extraction facilitated the analysis of qualitative data. We applied a predefined prompt that clearly bounded the task and required supporting quotations from the transcript. The model was therefore not asked to generate themes or conduct reflexive thematic analysis, and researchers could check its output against the supporting verbatim quotations. For two of the five groups, we compared the LLM-extracted information with that extracted through independent human analysis before using the LLM-assisted approach for the remaining groups. This comparison showed that the human-and LLM-extracted information largely overlapped.

The use of LLMs to assist qualitative research is therefore promising. However, tit also raises considerations regarding privacy and reliability. We de-identified the transcripts before submitting them to the LLM, a task that required manual effort. Additionally, we used Microsoft 365 Copilot under our university work account, whose enterprise data protection terms ensure that prompts and outputs are not used to train foundation models. Nevertheless, we recommend keeping the researcher in the loop, using specific prompts with clear boundaries, and requiring traceable evidence (such as verbatim quotes) to verify the quality of the LLM output. The interpretation of the findings should remain the responsibility of the researchers.

## 4.4 Limitations

Our work has several limitations. First, the workshop was hosted at a conference and participation was opt-in based, likely introducing selection bias toward individuals who already value co-creation and collaboration with LE experts, as observed in the responses to the pre-workshop questionnaire. Second, three research team members acting as facilitators may have introduced further bias, although they were excluded from the calculation of questionnaire scores in the main text. Including them in the calculations had only a minimal effect on the overall scores, as shown in Section 3.1.1. and Appendix D.

Third, the moderators' pre-workshop explanations of the objectives, the code of conduct, and the principles of co-creation may also have primed participants' responses. Note that the principles of co-creation (i.e., collaboration, inclusivity, transparency; see [38]) and code of conduct (i.e., work together, listen, build on each other's ideas, and propose out-of-the-box ideas) are all framed as inherently positive concepts, and therefore, participants may have felt prompted to rate the relevance of the activities more positively. Finally, although we attempted to create a safe and open environment for the LE experts to voice their opinions, power imbalances—for instance, between LE experts and their accompanying caregivers—may have influenced the group discussions, potentially contributing to the exceptionally high workshop ratings observed.

### 4.5 Recommendations for future research

The workshop tools and methodology provide a foundation for the early involvement of LE expertise in the development of neurorehabilitation technology. We recommend a series of adaptations for future iterations. First, we suggest including psychological and social dimensions of lived experience, such as confidence, self-efficacy, and aesthetics, to support a more holistic understanding of the LE experts' needs. This could be achieved by adding a section in the *Data Collection Notebook* summarizing what each LE expert values the most (e.g., independence in daily tasks over speed of recovery) and the emotions they express (e.g., feeling frustrated when a task highlights their disability), and by adding a section to the *Journey Map tool* that encourages groups to consider the psychological and social dimensions that may be linked to each step of the journey.

Second, future workshops may benefit from allocating more time to each phase of the Double Diamond model. The present workshop compressed the first three phases into two hours; future activities may consider longer sessions for each phase to allow deeper discussion, reflection, and convergence. Alternatively, multiple shorter sessions per phase might better accommodate LE experts' needs. Especially for individuals with more severe impairments, participation might be limited due to practical constraints, underrepresenting their experiences and needs. In such cases, the perspectives of caregivers and family members become essential. Furthermore, allocating more time to the workshop across multiple sessions could enable inclusion of the final Deliver phase of the Double Diamond model, allowing participants to test, evaluate, and iteratively refine the prototype with the LE expert(s).

Finally, a challenge remains in moving from ideas to implementation, a concern raised by one of the LE experts and reflected in the literature [62,63]. Although LE involvement supports the development of clinically relevant technologies [14–18], other relevant stakeholders should be included in the design process to ensure the technology reaches the final users. For example, clinical personnel's perceptions of technological complexity, along with limited time for patient setup, can restrict clinical uptake of neurorehabilitation technology [11,64], while commercial partners and insurers are essential stakeholders to support cost-effective development, reimbursement protocols, and viable business models [65]. Implementation should therefore be treated as an integral part of participatory design by involving LE experts, families, clinicians, researchers, companies, funders, insurers, and regulatory bodies from the start [66].

To achieve this level of multi-stakeholder collaboration, we encourage researchers to keep learning about and practicing participatory design methodologies [37,46,67,68]. Where possible, we also recommend incorporating transdisciplinary practice education into engineering and medical academic curricula to prepare future researchers and clinicians for this kind of collaboration. Finally, we feel that true connection and understanding of LE requires, as with most social relationships, sustained engagement, for example, by participating in community projects related to the disability.

## Acknowledgements

The authors sincerely thank Miguel Ángel Ferre Bañón for his assistance in the transcription of the audio recordings and Amy Bellitto and Alberto Garzás-Villar for their insightful contributions to the thematic analysis of the *Journey Maps*.

## Funding

This work was supported by the Health and Technology Convergence Alliance of TU Delft, Erasmus MC, and Erasmus University Rotterdam, the Dutch Research Council (NWO, VIDI Grant Nr. 18934), and the European Horizon 2022 Marie Skłodowska-Curie Actions (MSCA) Postdoctoral Fellowship under Grant 101106071.

## Conflicts of Interest

None declared.

## Data Availability

The dataset consisting of quantitative data (i.e., raw questionnaire responses) and the unfilled custom design tools used during the workshop are publicly available on Zenodo (10.5281/zenodo.22829440).

## Abbreviations

ADL: Activity of daily living

CEN: Centro Europeo de Neurociencias

COPM: Canadian Occupational Performance Measure

HLM: Hospital Los Madroños

ICNR: International Conference on NeuroRehabilitation

LE: lived experience

LLM: Large language model

## Appendix A

At the end of the presentations, workshop attendees had approximately ten minutes to get to know their teammates. To facilitate this process, they were introduced to the *Personal Profile Sheet*, a tool created by the research team, inviting participants to share their names, profession, and answer questions such as "*If you had to choose a different career, what would it be?*" (Figure A1). This icebreaker encouraged all participants to introduce themselves in a structured way and facilitated group interaction.

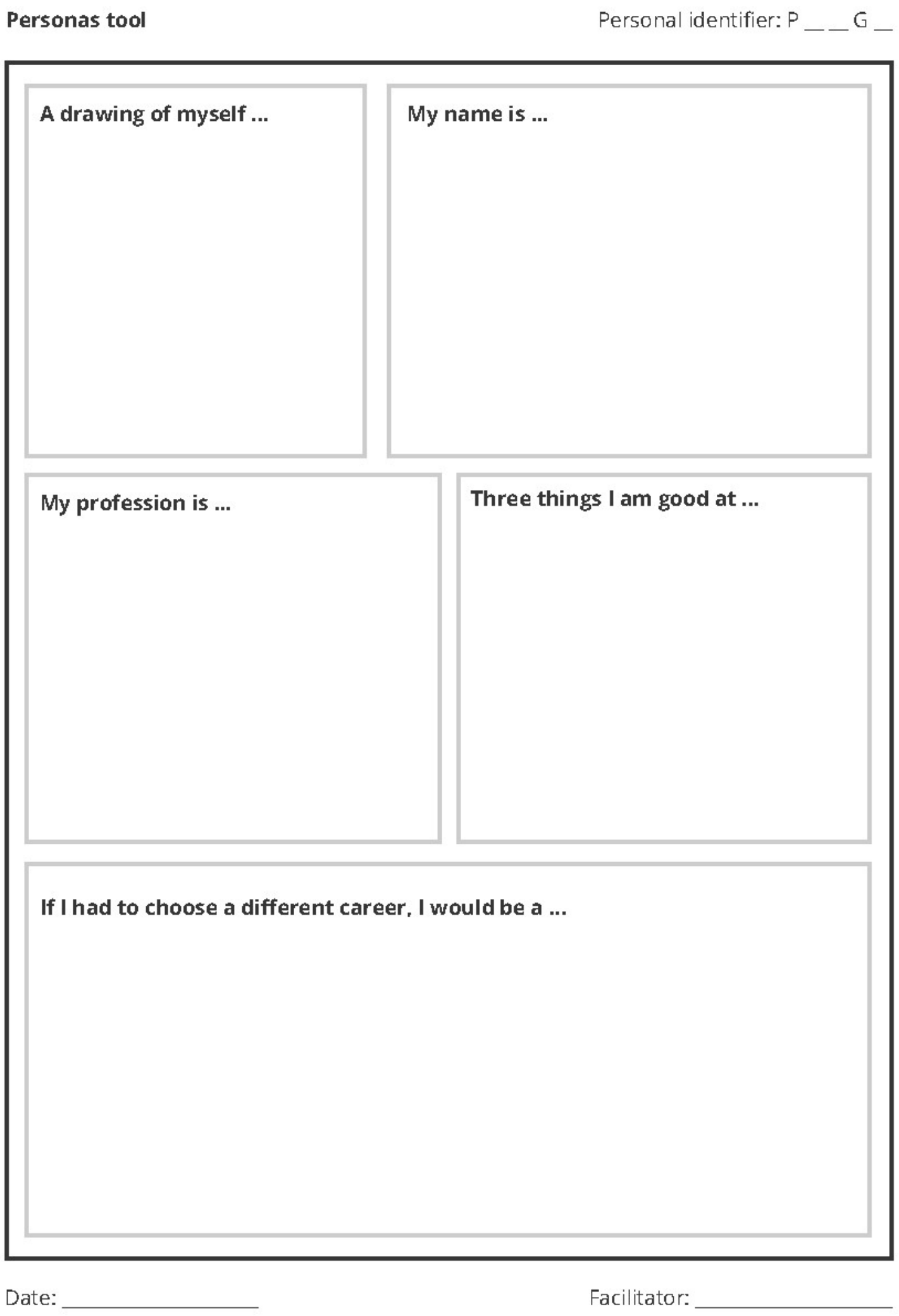

Personas tool

Personal identifier: P __ __ G __

A drawing of myself ...

My name is ...

My profession is ...

Three things I am good at ...

If I had to choose a different career, I would be a ...

Date: ____________

Facilitator: ____________

**Figure A1**. *Personal Profile Sheet* used during the workshop. This tool invited participants to complete the template individually and introduce themselves to their group members.

## Appendix B

**Table B1.** The seven open-ended questions in the workshop attendees' post-workshop questionnaire.

| N | Open question |
|---|---|
| 1 | What did you like the most about the workshop? |
| 2 | What did you like the least about the workshop? |
| 3 | What do you think about the active involvement of patients in the design process? |
| 4 | What other design methods or tools do you or your colleagues use or are aware of? |
| 5 | What suggestions do you have for improving the effectiveness of future workshops on the design process? |
| 6 | What additional resources or follow-up activities would help you implement the workshop's teachings more effectively? |
| 7 | What were your takeaways from the workshop? |

**Table B2.** The four open-ended questions in the LE experts' post-workshop questionnaire

| N | Open question |
|---|---|
| 1 | How would you describe your overall experience as an active participant in the technology design process? In what ways do you think your participation contributed to the design of the technology? |
| 2 | What aspects of the design process worked well for you, and what could be improved? What changes would you suggest for future design processes that involve end users (such as patients and caregivers)? |
| 3 | To what extent does the resulting solution meet your needs and expectations? |
| 4 | Did participating in the design process make you feel like you had more control over your recovery process? Why or why not? |

## Appendix C

**Table C1.** Outcome of the human and LLM extraction for Group A.

| Human extraction outcomes | LLM extraction outcomes |
|---|---|
| **Final product solution**<br><br>Something that helps a patient (suffering from a loss of proprioception on one side of their body: both arm and leg) to regain anticipatory postural response while riding a bike. This solution, which uses vibration signals, requires the user to move the affected limb, especially when s/he is losing balance. This solution must allow the user to re-learn how to perform this very task, not a motor that compensates for the loss. The solution must ensure the user maintains the necessary tension to balance while on the bike.<br><br>*"[..] the solution we discuss is something that helps him regain this anticipatory postural response [..] what we were thinking of was something that, for example, vibrates to signal that he needs to move the affected limb—both in the arm and in the leg, or, for example, when you're biking and you're losing your balance—and he's looking at a bird that's moving or looking somewhere else—then he doesn't realize that his balance is lost. So, something that signals that he is losing his balance. We were discussing whether what we need is a motor that readjusts the bike or something that signals him to do it, because what we want is for him to learn how to do it. It could be something that vibrates, or something that indicates he needs to reattain that tension to maintain his balance."* (Part_09; page 18)<br><br>**The 10 considerations** | **Final product solution**<br><br>The group's final solution was a bike-based sensory feedback/alert system for a stroke survivor with reduced proprioception on one side of the body. The idea was not to make the bike automatically correct itself, but to signal the patient—e.g., through vibration—when balance, limb position, or alignment is off, so he can learn to react and regain anticipatory postural control while cycling.<br><br>*"something that, for example, vibrates to signal that he needs to move the affected limb—both in the arm and in the leg" and "something that signals that he is losing his balance."*<br><br>*The presenter also clarified the design intent:*<br><br>*"whether what we need is a motor that readjusts the bike or something that signals him to do it, because what we want is for him to learn how to do it." These quotes are from (part_09)*<br><br>**The 10 considerations**<br><br>**Cycling as selected goal**<br><br>*"We've decided on the bike." (part_01)*<br><br>**Trunk control first**<br><br>*"before getting on the bike, the guy has to be able to stand up." (part_01)*<br><br>**Strength requirements** |
| **Balance reactions**<br><br>"*[..] because one thing is staying standing, and another is having balance reactions, equilibrium reactions [..]*". (Part_01; page 2)<br><br>**Pedaling with the affected leg**<br><br>"*[..] So the important thing is that one leg—maybe he can't pedal with the affected leg, but if he can do it with the other one and this one can do it [..] With one leg, you'll be fine. The problem is getting the other leg to match the rhythm of this one [..]*". (Part_01; page 2)<br><br>**Anticipatory posture reaction**<br><br>"*[..] if we don't want him to kill himself, we have to practice anticipatory postural adjustments [..] all of this is so that I'm able to pedal, and when I want to stop pedaling, I can react by pushing off the ground without falling [..]*" (Part_02; page 2)<br><br>**More than a motor issue**<br><br>"*[..] there are people who physically have the ability, but then their comprehension has been affected, so the patient doesn't remember the steps they have to take [..] the patient has to get used to going alone and goes carefully because their motor and mental control aren't the same—they're affected [..]*" (Part_01 & 02; pages 2-3)<br><br>**Running behind them**<br><br>"*[..] you go out on the street and take it slowly [..] and we run behind them and stuff [..] what we do here is supervise—on the uphills and downhills [..] If we see they're not stable here, we won't go there [..]*" (Part_03; page 5)<br><br>**That scary balance** | *"They have to have strength." (part_01)*<br><br>**Standing balance**<br><br>*"one thing is staying standing, and another is having balance reactions" (part_01)*<br><br>**Lower-limb coordination**<br><br>*"So it's coordination within the same side and between the two legs" (part_01)*<br><br>**Anticipatory reactions**<br><br>*"we have to practice anticipatory postural adjustments" (part_01)*<br><br>**Cognitive sequencing**<br><br>*"the patient doesn't remember the steps they have to take. So they get on the bike and don't know what to do." (part_01)*<br><br>**Real-bike practice**<br><br>*"Practice, wheel, rear, bike, on rollers." (part_02)*<br><br>**Proprioception loss**<br><br>*"there's a lack of confidence, which in Spanish is... Confianza" and "a lack of proprioception" (part_08)*<br><br>**Feedback over automation**<br><br>*"what we want is for him to learn how to do it" (part_09)* |
| *"[..] balance scares me the most. I get the feeling that I wouldn't be able to... to hold you up [..]"*. (Part_06; page 12)<br><br>**Keep (the bike) moving**<br><br>*"[..] usually what's hardest when you're learning to ride a bike is getting the bike moving [..] the feeling I have right now would be coordinating, grabbing the handlebars, coordinating, moving my leg, and at the same time trying not to fall."* (Part_06; page 12)<br><br>**Security also has its dangers**<br><br>*"[..] in your current situation, it would actually be easier if you had the clip—or rather, to remove the clip to release it—it makes me feel less secure [..] there are lots of people who [..] ride without cleats for the safety of being able to release their foot [..]it also has its dangers. Because your foot can get caught."* (Part_07; page 14)<br><br>**Building up confidence**<br><br>*"[..] you don't feel confident riding your bike on the street, right? He doesn't feel his leg or his arm unless he looks at them. Sure, you right now, without looking at your feet, do you know how they're positioned? I mean, a lack of proprioception [..] He won't be able to keep track of his leg or his arm [..]"* (Part_08; page 15)<br><br>**Not feeling it**<br><br>*"[..] he doesn't control his body, effectively, because he doesn't quite feel it [..] If they don't feel it, you don't move [..] there's no coordination either [..]"* (Part_08; page 16) | |
| **The 5 patients' considerations** | **The 5 patients' considerations** |
| **Good Spirit and new perspectives**<br><br>"*[..] the first two days this happened to me, I thought about selling the motorcycles, the bikes, and everything, but by the third day I wasn't even thinking about selling them anymore. The first two days, even though I wasn't feeling bad—I think I was in good spirits—but anyway, you think about it in terms of, "This should have that there," and then once it's actually up and running, I see it from a different perspective [..]*" (Part_06; page 12)<br><br>**Being scared of falling**<br><br>"*[..] balance scares me the most. I get the feeling that I wouldn't be able to... to hold you up [..]*". (Part_06; page 12)<br><br>**Do it real**<br><br>"*[..] It feels more natural to me. Seeing things [..]*". (Part_06; page 12)<br><br>**Having the right conditions**<br><br>"*[..] not because I prefer the road bike, but because on a road with no traffic just for me, well... There's less friction; it moves more easily [..]*" (Part_07; page 13)<br><br>**Confidence in your body** | **Balance fear**<br><br>*"Balance—balance scares me the most." (part_06)*<br><br>**Arm and leg limitation**<br><br>*"My leg and my arm." (part_06)*<br><br>**Coordination difficulty**<br><br>*"coordinating, grabbing the handlebars, coordinating, moving my leg, and at the same time trying not to fall." (part_06)*<br><br>**Outdoor cycling preference**<br><br>*"It feels more natural to me. Seeing things." (part_06)*<br><br>**Traffic safety concern**<br><br>*"you don't know if the person behind you is coming up on you safely or dangerously." (part_07)* |
| "*[..] It's because you don't have the same confidence in your body that you had before.*"<br>(Part_07; page 14) | |

**Table C2.** Outcome of the LLM extraction for Group B.

**Final product solution**

The group's final product solution was a shower-assist device combining body support with an automated rotating washing/foaming mechanism. The patient could lean/hold on for balance, while a spinning up-and-down "bubble-making" / washing element helps clean the neglected side of the body; afterward, the shower rinses everything off. In the final presentation, the group described it as:

"*a bubble-making machine here. So he could lean over first and hold on, and then this one spins up and down… and then he turns on the shower and washes everything off—that's our solution.*" (Part_09)

**The 10 considerations**

**Whole shower journey**

"It's like the whole journey… to go into the shower, take a shower, get dressed, and then leave." (Part_01)

**Shower over bathtub**

"maybe a shower would be best and better than a bathtub." (Part_01)

**One-handed operation**

"you have to remember he only has one hand." (Part_01)

**Temperature control**

"If there's a change in the water… cold water comes out instead of hot—I can't even imagine the screams." (Part_06)

**Objects fixed in place**

"it's very important that everything is in its place every day: the faucet, the shampoo, or the gel." (Part_06)

**Standing, not sitting**

"For me, it's very important to shower standing up, not sitting down." (Part_06)

**Need for balance**

"I always have something to hold onto; I hold on, I'm safe, I'm fine." (Part_06)

| |
|---|
| **Neglect of left side**<br><br>"Maybe washing only half my head because of the neglect." (Part_06)<br><br>**Hypersensitivity to drops**<br><br>"A single drop—if a drop falls, it's like fire, fire, fire that shoots right through you." (Part_07)<br><br>**Autonomy and confidence**<br><br>"I shower, all by myself. I don't let ******* take over—I do it myself." (Part_08) |
| **The 5 patients' considerations**<br><br>**Avoid temperature changes**<br><br>"For me, showering is a very unpleasant process. Because of hypersensitivity. To the temperature." (Part_06)<br><br>**Keep setup unchanged**<br><br>"everything is set up so there aren't any sudden problems." (Part_06)<br><br>**Stand while showering**<br><br>"For me, it's very important to shower standing up, not sitting down." (Part_06)<br><br>**Support left-side awareness**<br><br>"The left side doesn't exist; he has to think about that so much." (Part_08)<br><br>**Preserve independence**<br><br>"Always on my own, perfect. I shower, all by myself." (Part_08) |

**Table C3.** Outcome of the human analysis and LLM analysis for Group C.

| | |
|---|---|
| **Final product solution**<br><br>A splint that is stylish but with integrated stimulation so the hand can grasp an object. The activation would be through voice control or ideally using surface EMG of the muscle signal.<br><br>"*so we thought of a sort of splint so that could be also nice like this one but we've integrated the stimulation so that she she could have stimulation on the hand to let's say grasp an object and we discussed also how this could be activated so in the beginning we thought maybe with a button or with a Bluetooth or something but of course if one hand is engaged in other activities we'd also discuss potentially voice control but maybe ideally would be with some surface electromyography*"<br><br>**The 10 considerations**<br><br>**Low sensation in arm**<br><br>"*I felt it, but it's very faint [..] Especially from the elbow below*"<br><br>**Bimanual manipulation**<br><br>"*Being able to take a tray out of the oven, which right now is really heavy for me with just one hand*"<br><br>**Support weight** | **Final product solution**<br><br>The group's final product concept was a stylish wearable wrist/hand splint or brace with integrated stimulation/electrodes, designed to help the patient's affected left hand grasp objects and stabilize the wrist during cooking and other daily activities. The preferred activation method evolved from simple options such as a button or Bluetooth toward voice control or, ideally, surface electromyography / muscle-signal activation, because the patient still has some residual movement and muscle signal.<br><br>*"we thought of a splint like this one, but we've integrated stimulation so she can receive stimulation on her hand to grasp an object, and we also discussed how this could be activated." (Part_6)*<br><br>*"Ideally, though, it would be with surface electromyography so that, since she has some movement, she has a muscle signal to activate the holding function and also to stabilize the wrist—that was the main idea." (Part_6)*<br><br>**The 10 considerations**<br><br>**Cooking as target activity**<br><br>*"the goal then was to cook with the left arm" (Part 3)*<br><br>**Reaching objects** |
| *“[..] but also a certain level of strength to support the weight ”*<br><br>**Not be bothersome**<br><br>*“So my goal with the arm [..] is for it to bother me as little as possible”*<br><br>**Functionality in ADL**<br><br>*“[..] for it to be as functional as possible for getting dressed*"<br><br>**Aesthetics**<br><br>*“[..] for it to look good, and for it not to feel so heavy"*<br><br>**Remote activation**<br><br>*“[..] that I could activate remotely with this part, so that it would provide some kind of support”*<br><br>**Hand grasping**<br><br>*“The grip is what’s hardest for you”*<br><br>**Wrist support**<br><br>*“[..] something for the wrist, to be able to work on the wrist ... so my hand doesn’t drop so much”* | *“the first step we’d need to work on—or where we’d begin the activity—is reaching” (Part 3)*<br><br>**Grasping objects**<br><br>*“which would be the next step: grabbing the object.” (Part 3)*<br><br>**Force and movement control**<br><br>*“coordination to control both the force and the movement to adapt to the different surfaces or densities” (Part 3)*<br><br>**Bimanual support role**<br><br>*“your left arm would support the activities without performing the main function.” (Part 3)*<br><br>**Transporting objects**<br><br>*“Okay, because then we’d be on to the next step: transporting objects.” (Part 3)*<br><br>**Grip as hardest problem**<br><br>*“The grip of the hand. The grip of the hand.” (Part 4)*<br><br>**Wrist stabilization needed**<br><br>*“something for the wrist, to be able to work on the wrist.” (Part 3)*<br><br>**Aesthetic wearability matters**<br><br>*“the aesthetic aspect is important to her as well.” (Part_6)*<br><br>**Hands-free activation preferred** |
| **Gross motor control**<br><br>*"[..] but hold on a little bit, using minimal amount of force [..] Yes, more on a gross motor level".* | *"if one hand is occupied with other activities, we also discussed the possibility of voice control." (Part_6)* |
| **The 5 patients' considerations**<br><br>**Part of the outfit**<br><br>*"I already wear a wristband. I customize them and design them to match my outfit"*<br><br>**Supporting the dominant hand**<br><br>*"It's more about helping and collaborating with the hand, with the dominant side"*<br><br>**Integrating the left arm**<br><br>*"She'd like to be able to integrate it more, use it more in her day-to-day life – in the different activities she does now."*<br><br>**Given up on hand**<br><br>*"We're using the wrist, but the hand, for example, I've given up on that – I've given it up because it's impossible"*<br><br>**Aesthetics** | **The 5 patients' considerations**<br><br>**Arm should bother less**<br><br>*"my goal with the arm, even though we have this exercise, is for it to bother me as little as possible" (Part 3)*<br><br>**Dressing functionality**<br><br>*"for it to be as functional as possible for getting dressed" (Part 3)*<br><br>**Wrist should not drop**<br><br>*"So my hand doesn't drop so much." (Part 3)*<br><br>**Grip is difficult**<br><br>*"Grip is hard for me" (Part 3)*<br><br>**Stylish, non-bulky device**<br><br>*"I customize them and design them to match my outfit" (Part_5)* |
| *“[..] for it to look good, and for it not to feel so heavy”.* | |

**Table C4.** Outcome of the LLM analysis for Group D.

**Final product solution**

The group designed a pliers-like pill/blister-pack opener that lets the patient place a blister pack into the tool and press using the whole hand, converting the task from painful fine fingertip movements into easier gross hand movements. The patient also suggested adding a small box/cup attached to the pliers so the pill does not shoot away or fall when released.

*"We came up with a kind of pliers where you can place the blisters, and then, using the whole hand, something pushes the blister and removes the pill." (part_4)*

*"He also suggests securing a box in the pliers to hold the pill in place while pressing." (part_5)*

**The 10 considerations**

**Opening blister is hardest**

*"Everything we've talked about has been the blister pack. The blister pack. The blister pack—that's the one with the most problems." (part_2)*

**Focus on opening**

*"I'd almost focus on, rather than taking it out, opening it." (part_2)*

**Plastic/aluminum stiffness**

*"if it's plastic or aluminum foil, which is a little stiff or something, or it doesn't fit the shape well." (part_2)*

**Weak fingertip pinch**

*"The strength to squeeze here, at the tips of the fingers, here with the fingers—I do have some differences there." (part_1)*

**Sensory loss causes pain**

*"this lack of sensation causes him persistent pain, so while he can do it, he does so with great discomfort." (part_4)*

**Pain lasts afterward**

*"the pain lasts a long time afterward because I still feel the tingling in my fingers." (part_2)*

| |
|---|
| **Fine movement problematic**<br><br>*"The main problem is the delicate movement of removing it because it causes a persistent sensation of pain" (part_2)*<br><br>**Use gross movement**<br><br>*"the main way to help him do this properly and comfortably was to shift the activity from fine motor skills to gross motor movements." (part_4)*<br><br>**Whole-hand pressing**<br><br>*"But press with your hand. With your hand, yes. Instead of with your fingers. With a broad motion." (part_3)*<br><br>**Catch falling pill**<br><br>*"the part of the glass he proposed is... Yes, yes... that the pills shoot out." (part_3)* |
| **The 5 patients' considerations**<br><br>**Pre-cut blisters prick**<br><br>*"when I go to open it I even prick myself on the sharp edge" (part_1)*<br><br>**Full blister easier**<br><br>*"When the blister pack is full, it's easy because you can push a little harder or slide it down" (part_1)*<br><br>**Small folds hard**<br><br>*"the tear, when the fold is small. When the fold is small, right?" (part_1)*<br><br>**Small pills difficult**<br><br>*"If it's too small, it has trouble; if it's too small, it's a pain in the ass." (part_1)*<br><br>**Add collection box**<br><br>*"he also suggests securing a box in the pliers to hold the pill in place while pressing" (part_5)* |

**Table C5.** Outcome of the LLM analysis for Group E.

**Final product solution**

The group's final product solution was an adapted one-handed nail clipper: a nail cutter mounted on a stable table platform/base, with adjustable cutter positioning and an extended press-down lever/plate so the patient can cut nails using the functional right hand while keeping direct visual control and reducing the need to use the affected left hand. In the final presentation, the group described it as:

*"this cutter should have a kind of platform to stabilize it on a table," with the platform allowing the cutter position "to be adjusted slightly," plus "a knife to extend the part you press down on the cutter" so "you can see what you're doing while you're doing it."*

**The 10 considerations**

**Right hand usable**

*"she can use her right hand properly" (part_1)*

**Left hand immobile**

*"she has no movement in her left hand" (part_1)*

**Right nails difficult**

*"she can cut her left fingernails, but she can't do it on her right hand." (part_1)*

**Stable positioning needed**

*"placing it in a stable position so that you don't need to—maybe you don't need to hold it." (part_1)*

**Alternative pressing method**

*"you have to find another way, you have to press the cutter" (part_2)*

**Avoid dual-task complexity**

*"I'm not yet able to... press down and hit it with my head at the same time." (part_3)*

**Wider pressing area**

*"the part of the nail clipper here could be just a little wider" (part_4)*

**Precision and visibility**

| |
|---|
| *"it might be better to see what you're doing and have something that can be precise." (part_4)*<br><br>**Filing too tiring**<br><br>*"Besides taking a long time, I end up getting tired of the tasks involved" (part_4)*<br><br>**Longer lever helps**<br><br>*"if it's longer, you also have to press down a little, so you're more precise too." (part_5)* |
| **The 5 patients' considerations**<br><br>**Step order matters**<br><br>*"First I put my hand down, and then I take the nail clipper." (part_3)*<br><br>**Thumb has control**<br><br>*"Because I have strength in my thumb. And I can move it, yes." (part_3)*<br><br>**Attention limits multitasking**<br><br>*"I have attention problems." (part_3)*<br><br>**Tool must be wider**<br><br>*"if it's too thin, there's a higher chance it'll get stuck." (part_4)*<br><br>**Filing causes fatigue**<br><br>*"I end up getting tired of the tasks involved, so I end up giving up" (part_4)* |

## Appendix D

The mean and standard deviations of the responses to the pre- and post-workshop questionnaires were recalculated including the four co-authors' data in Tables D1, D2 and D3.

**Table D1**. Workshop attendees' responses to five pre-workshop items. [a]

| **Statement** | **Pre-workshop score**<br>**Mean (*SD*)** |
|---|---|
| How much experience do you have with interdisciplinary collaboration? (1 = No experience, 5 = Expert level) | 4.6 (0.7) |
| I have experience working with end-users (e.g., patients and caregivers). (1 = Not at all and 5 = Very much so) | 4.2 (1.1) |
| Interactions with end-users (e.g., patients and caregivers), both personally and professionally, can improve my approach to designing novel technologies and solutions. (1 = Not at all and 5 = Very much so) | 4.6 (0.6) |
| Working with end-users (e.g., patients and caregivers) allows me to better empathize with them and increase my understanding of their needs and challenges. (1 = Not at all and 5 = Very much so) | 4.8 (0.4) |
| I enjoy working with end-users (e.g., patients and caregivers). (1 = Not at all and 5 = Very much so) | 4.5 (1.1) |

[a]*Note*. *n* = 15 for the first item and *n* = 17 for the remaining four items.

**Table D2**. Workshop attendees' responses to the five questions administered both pre- and post-workshop. Anchors per question are shown in the table. [a]

| Question | Pre-workshop score<br>Mean (*SD*) | Post-workshop score<br>Mean (*SD*) |
|---|---|---|
| How much experience do you have with co-creation workshops?<br>(1 = No experience, 5 = Expert level) | 2.1 (1.2) | 2.7 (1.4) |
| How much do you value co-creation methods?<br>(1 = Do not value, 5 = Highly value) | 4.4 (0.9) | 4.7 (0.5) |
| How confident are you in the application of co-creation methods?<br>(1 = Not confident, 5 = Very confident) | 3.2 (1.1) | 4.1 (0.9) |
| How much do you value interdisciplinary collaboration?<br>(1 = Do not value, 5 = Highly value) | 4.9 (0.5) | 4.9 (0.4) |
| How confident are you in your ability to collaborate across disciplines?<br>(1 = Not confident, 5 = Very confident) | 4.2 (1.0) | 4.4 (0.9) |

[a]*Note.* Pre-workshop $n$ = 15 for the first four items and n = 17 for the fifth; post-workshop $n$ = 15.

**Table D3**. Workshop attendees' agreement with statements post-workshop. 1 = Not at all, 5 = Very much so. [a]

| **Statement** | **Post-workshop score**<br>**Mean (*SD*)** |
|---|---|
| I felt actively engaged in this workshop | 4.7 (0.6) |
| My expertise contributed significantly to the success of the group | 4.1 (0.7) |
| I am likely to apply the methods and tools learned in this workshop to future design projects. | 4.3 (0.7) |
| This workshop helped me identify strengths and weaknesses in my current design process. | 4.1 (1.0) |
| End-users (e.g., patients and caregivers) participation enhanced the design process. | 4.9 (0.3) |
| Collaborating with end-users (e.g., patients and caregivers) was a valuable experience. | 4.9 (0.3) |
| Empathizing with end-users (e.g., patients and caregivers) is essential for designing solutions that meet their needs. | 5.0 (0.0) |
| Understanding end-users (e.g., patients and caregivers) needs is crucial for defining design requirements | 5.0 (0.0) |
| Working directly with end-users (e.g., patients and caregivers) could be effective for understanding their needs. | 4.9 (0.3) |
| The involvement of end-users (e.g., patients and caregivers) made me reconsider my approach to design. | 4.5 (0.8) |
| Working with end-users (e.g., patients and caregivers) allowed me to better empathize with them and increased my understanding of their needs and challenges | 4.9 (0.3) |
| I plan to actively involve end-users (e.g., patients and caregivers) in future design projects. | 4.8 (0.4) |
| Interactions with end-users (e.g., patients and caregivers) improved my approach to novel technologies and solutions. | 4.6 (0.6) |
| I believe co-creation could be an effective method for designing meaningful solutions. | 4.7 (0.5) |

[a]*Note*. *n* = 15.